\documentclass[12pt]{article}
\usepackage{graphicx}
\usepackage{amssymb,amsmath,amsfonts,palatino,amsthm}
\usepackage{amssymb}
\usepackage{epstopdf}
\newcommand{\f}{\begin{equation}}
\newcommand{\ff}{\end{equation}}

\begin{document}

\title{Quantum reference frames and triality}
\author{Lee Smolin\thanks{lsmolin@perimeterinstitute.ca} 
\\
\\
Perimeter Institute for Theoretical Physics,\\
31 Caroline Street North, Waterloo, Ontario N2J 2Y5, Canada\\
and\\
Department of Physics and Astronomy, University of Waterloo\\
and\\
Department of Philosophy, University of Toronto}
\date{\today}
\maketitle

\begin{abstract}

In a background independent theory without boundary, physical observables may be defined with respect to dynamical reference systems.  However, I argue here that there may be a symmetry that exchanges the degrees of freedom of the physical frame of reference with the other degrees of freedom which are measured relative to that frame.   This symmetry expresses
the fact that the choice of frame of reference is arbitrary, but the same laws apply to all, including observer and observed..

It is then suggested that, in a canonical description, this leads to an extension of the Born duality, which exchanges coordinate and momentum variables to a triality that mixes both with the temporal reference frame.   This can also be expressed by extending $2n$ dimensional symplectic geometry to a $d= 2n+1$ dimensional geometry with a cubic invariant.  The choice of a temporal reference frame breaks the triality of the cubic invariant to the duality represented by the canonical two form.

We discover that a very elegant way to display this structure which encompasses both classical and quantum mechanics, is in terms
of matrix models based on a cubic action.  There we see explicitly in either case how a spontaneous symmetry breaking
leads to the emergence of a temporal reference frame.

Note::  This article draws partly on \cite{Universal} and in part
supersedes \cite{triality}.
\end{abstract}

\newpage

\tableofcontents



\section{Introduction: triality and dynamical reference frames}

One of the most fundamental principles activating our search for a common completion of general relativity and quantum mechanics
is background independence.  This principle asserts that the laws of physics not depend on any, fixed, non-dynamical background structures\cite{TN,3R,TR,SURT}.  Background independence underlies all our modern understanding of the fundamental forces as its avatars in field theory are diffeomorphism invariance and local gauge 
invariance

Another consequence of background independence is that the common completion of quantum mechanics and general relativity must be a cosmological theory.    In \cite{SURT}  Mangabeira Unger and I make a detailed analysis of how a theory of the whole universe as a closed system must differ from a theory of a portion of the universe.   This is also argued in 
\cite{TN,3R,TR}.

In particular a theory of the whole universe contains its observers and frames of reference as dynamical subsystems.   As Einstein
taught us,   any complete description of a physical observable contains, implicitly at least, a description of how a frame of reference,
which gives a precise meaning to that observable, is
established in each concrete application to a physical system. Most importantly, any theory that ascribes times to physical events, must contain a dynamical description of an internal clock.   Similarly, any 
use of location or velocity must begin as a measurement of relative position-between the system under study and a physical
system that is defined as a frame of reference.

It is very important to understand that fundamentally, nothing-except a certain very minimal level of complexity, distinguishes the  physical subsystems of the universe that we choose to play the  roles of frame of reference and/or observer, from other subsystems, which we use the frame of reference to describe.  
{\it Hence, any description that employs one subsystem as the frame of reference, whose clock readings and rulers define physically
meaningful quantities, breaks a symmetry.}

That symmetry mixes up observer and observed, as it expresses the obviously true fact that both the observer or frame of reference and the physical systems whose description they anchor are physical subsystems of the universe.  Consequently, so long as each has sufficient complexity,  it must be possible to give a description in which the 
two systems trade places-and what was the physical system is now the frame of reference, and  vice versa.

This implies that there must be two novel structural elements:  first, a symmetrical description of the system in which the degrees of freedom usually described are mixed up with degrees of freedom which, when frozen, create a reference system.
Second, this symmetry must be spontaneously broken, in a way that freezes and decouples those degrees of freedom
that will serve as the reference systems.


It follows directly from this that if there is already a symmetry amongst a number of the particles of the system, say a
permutation symmetry, $P_N$ or a continuous symmetry $SO(N)$, including the frame of reference might increase this to an
$S_{N+1}$ or $SO(N+1)$.   In the canonical formalism this might extend an $sp(n)$ symmetry
of symplectic geometry to an $SO(d+1,d)$ symmetry.  

We should also note the subtle behaviour of degrees of freedom attributed to physical reference frames.  To serve their purpose, they must be very weakly coupled to the other physical
degrees of freedom, hence they must be close to a limit in which some decoupling 
parameter, $\lambda $ is taken to zero.  But one cannot take the limit in which 
$\lambda =0$ lest they decouple completely.  Often one is interested then in only
the linear, response of the measuring instruments to the physical degrees of freedom
they give access to.

This paper continues and brings together, several previous lines of research, on the 
basis of the new perspective offered by the investigations into dynamical and
quantum frames of reference\cite{QRF0}-\cite{QRF4}\cite{QRF-QG}.  These include the proposal to extend dualities to
trialities put forward in \cite{triality}.    A model for this extension is found in the study of a special kind of matrix model, cubic in the matrices\cite{cmm}-\cite{Universal}.  
These are models which have a diverse set of avators, which may emerge from them, including topological field theories, gauge theories and theories of gravity\cite{Universal}.

In this paper, we study first the most elementary cases, in which the reference degree of freedom is only a time parameter,  so the gauge
symmetry which enforces its redundency is time reparametrization.   This is spontaneously broken, freezing the choice of time
variables, which may then be identified as the reading of an internal clock.  

We next locate the symmetry whose breaking gives rise to the separation of a frame of reference which includes a clock, 
as an augmentation of the fundamental Born duality to a triality.  The extra, third degree of freedom is a gauge parameter that under
symmetry breaking to distinguish it, it becomes an operator that measures the time rate of change of other operators..


 We will see in detail how this works in both classical and quantum mechanics.

In the classical case this transformation amongst frames of reference is found to be an element of a group of transformations-
such as the Galilean or Poincare groups.   But the quantum case is believed to be considerably more intricate, because the observer and system may be entangled or in superposition of states in which the frame of reference has definite values.

Here, I propose a general solution to this cluster of problems based on the 
following schema (or principles\footnote{I want  to emphasize that this is not the only approach to these 
issues, see  for example \cite{QRF-QG} or \cite{QRF0}-\cite{QRF4}.}.)

1)  Both classical and quantum dynamics are built on dualities.  In classical mechanics this is the 
Born duality between momentum and position\cite{laurent-string}.
In quantum mechanics we also have a form of Born duality; and we have as well the duality  between bra and ket: or in other words between preparation and measurement.  

2)   In both cases the duality pivots on a fixed, static point, which represents a fixed frame of reference.   These are the clock, (the choice of which is also represented in the choice of symplectic
structure (in classical physics),  and the hermitian inner product of Hilbert space.  This is no surprise, as we know the 
symplectic structure and the Hilbert space inner product are closely tied to each other.  

3)  In order to realize the symmetry that trades physical system and frame of reference, we must enlarge the duality to a triality, as suggested in \cite{triality},
whose
third element is precisely the symplectic structure or Hilbert space norm that gives meaning to  the basic observations (which are always of
relative quantities.).   The classical triality is
\f
\ \ \ \ \ x^a \leftrightarrow \ \ \  p_b \leftrightarrow \ \ \ \ \ \ \   \frac{d}{dt}     \ \ \ \ \ \ \ 
\ff
The quantum version is
\f
\ \ \ \ \   <\Psi |     \leftrightarrow \ \ \  | \Phi >     \leftrightarrow        \dagger   \ \ \ \ \ \ \    
\ff

I also describe below, the same idea expressed in terms of an extension of symplectic geometry, in which a $2n$ dimensional
phase space, $\Gamma$ is augmented by the addition of an explicit clock variable, to a $d=2n+1$ dimensional
space, $\tau$.  The quadratic and antisymmetric Poisson structure, $\Omega$,  on $\Gamma$ is augmented to a triple product, $\Phi$, 
antisymmetric, on $\tau$.  $\Phi$ is represented by a three-form on $\tau$, and  we can see  explicitly how its symmetry is broken by
the choice of a time variable, reducing the three form, $\Phi$  to the symplectic two-form   $\Omega$.

In this paper I present a model for how physical reference frames emerge from a spontaneous breaking of this triality symmetry
in both classical and quantum physics.   The model is a particular matrix model\cite{m}-\cite{others}, by which is meant a model whose degrees
of freedom are described as large matrices.  The meaning of large is that we will often want 
to take limits in which the dimension of the matrices are taken to infinity.
This, as we shall see, is necessary for quantum physics (ie the Heisenberg algebra) to fully emerge.

There is a vast literature on such models; the ones I will be interested in here have the feature that their actions contain or are
dominated by the simplest possible non-linear 
dynamics\cite{cmm}-\cite{Universal}:
\f
S^3 = Tr M^3
\label{S3}
\ff
where $M$ is an traceless, Hermitian $N \times N$ matrix.
These  turn out to model well the triality I spoke about above, that underlines the emergence of a quantum reference frame. 

To see how triality emerge, let $N=2M$ and, recalling that $M$ is required to be trace-free,  consider the following tensor product
decomposition of $M$
\f
M=\sum_J \tau^J  \otimes M_J
\ff
where the three $\tau^I, \ \ \ \ I= 1,2,3$ are the three trace-free Pauli matrices and
$M_J$ are three $M \times M$ matrices.   Then the action (\ref{S3}) is
\f
S^{3} = \epsilon^{IJK}   Tr \left ( M_I M_J M_K  \right )
\ff

We will see that a very simple matrix system such as the above, can give models of both
 non-relativitstic and relativistic  dynamics, classical and quantum, in a way that illustrates the role on internal reference systems and their
 symmetries. We will examine this in the next section, 2.   Remarkably, as shown in \cite{Universal}, they also may describe versions of string theory and topological field theories, including Chern-Simons theories.
 
 Among the many papers which develop these techniques, some of hte clearest are two  review papers of Washington Taylor
 \cite{Taylor}.  These give a review of the most important tool we will need.
  A summary is also provided in the second Chapter of \cite{Universal}

A final remark:  To motivate the truncations we study we may recall the idea of noiseless or decoherence free subspaces, from quantum information theory. In these systems,  persistent, physical degrees of freedom are brought into existence by splitting the system into subspaces, in a way that introduces
symmetries in the interactions of those subspaces with their environments\cite{nfs}.  Indeed it has been argued that in some cases emergent gauge symmetries can be understood as arising from noiseless subsystems\cite{nfs-gauge}, and that this may be the origin of physical degrees of freedom in background 
independent systems\cite{fot-dav}.  


Finally,   some of the text that follows are revisions of text from earlier papers
where these models were presented\cite{triality,Universal} 
but without the unifying theme of this paper.

\section{Examples of extending dualities to trialities: classical theories}

\begin{enumerate}

\item{}The triality behind classical mechanics.

It is worth emphasizing that the equations of motion in physics have a particular structure which distinguishes them from other systems that science studies that evolve in time.  

This is the fact that the dynamical variables come in pairs, which we call configurations, $x^a$
and momenta, $p_a$ that satisfy Poisson bracket relations
\f
\{ x^a, p_b \} = \delta^a_b
\ff

which have equations of motion of the characteristic form
\f
\dot{x}^a = q^{ab} p_b, \ \ \ \ \ \ \dot{p}_a = -\nabla_a   V(x)
\ff
This canonical structure is imparted by the first term in the action
\f
S^{symp}  =\oint ds  p_a (s) \frac{dx^a (s)}{ds} 
\label{symp0}
\ff
From this we see the Born duality
\f
x^a \leftrightarrow p_b
\ff

Now let's look at this structure from a background independent point of view (notice that there is no metric in (\ref{symp0}))
and we see that time appears to be a background structure.  However, note that the symplectic action (\ref{symp0})
is invariant under reparameterizations of the time coordinate,
\f
s \rightarrow s^\prime = f(s)
\ff
Hence, there is a hint of a larger triality symmetry
\f
x^a \leftrightarrow p_b \leftrightarrow  \ \frac{d}{ds}
\label{triality17}
\ff
To bring this out, we have to represent the system in terms of three dynamical objects, $M_I $, where $I=0,1,2$ are
\f
x^a  (s) \rightarrow  M_1 ,  p_b  (s) \rightarrow  M_2, \\ \ \  \frac{d}{ds} \rightarrow M_0
 \ff
In the next section,  we will see in detail that this can be realized by an algebra of three $N \times N$ matrices, in a limit, $N\rightarrow \infty$,     
chosen so that
\f
S^{N}=
Tr M_2  M_0 M_1 =\oint ds Tr_{M}  p_a (s) \frac{dx^a (s)}{ds} 
\ff

Hence the hint of a triality (\ref{triality17}) is realized as both an $S_3$ permutation symmetry and an $SO(3)$ gauge symmetry of the matrix theory action,

\f
S^{N}= \epsilon^{IJK}  \frac{1}{6} Tr M_I  M_J M_K 
\label{cmm}
\ff
where $I=0,1,2$.


\item{}  Triality from an extension of symplectic geometry.

This is a slightly more formal version of how this works in classical mechanics.  We describe the phase space
${\cal P} = ( q^a,  p_b  )$ as a  $2d$ dimensional symplectic 
manifold whose coordinates can be gotten if we use {\it the metric on momentum space}, $h^{ab}$ to raise the one forms
$p_a$ to vectors   
\f
p^a= h^{ab} p_b.
\ff
We then define coordinates
\f
X^I = ( x^a, p^b )
\ff
with $I=1....2d$.       

We can then define the symplectic two form as
\f
\Omega= \Omega^{JK} dX_J \wedge d X_K
\ff
Its properties are that 
\f
\Omega^{JK} = - \Omega^{KJ} ,   \ \ \   d\Omega_{IJ} =0
\ff

In terms of it, the Poisson bracket is       
\f
\{ f(X), g(X) \}   = \Omega^{IJ}\frac{df}{dX^I} \frac{dg}{dX^J} = \frac{df}{dq^a} \frac{dg}{dp_a} -\frac{df}{dp_b} \frac{dg}{dq^b}
\ff

The symplectic action (\ref{symp0}) is then
\f
\Gamma= \int ds X^J(s) \Omega_{JK} \frac{d X^K(s)}{ds}   = \int ds p_a (s) \frac{d q^a}{ds} =
\ff

This has symplectic symmetry $Sp(d)$ as well as a discrete "Born duality" under
\f
q^a    \leftrightarrow    p^a
\ff

To extend this duality to a triality we introduce one more dimension, $x^0$, giving us $SO(d+1,d)$ invariant manifold, $\cal Q$.
Its coordinates are 
\f
X^\mu = (x^0, X^I ) = (x^0, q^a , p^b    )
\ff

We  extend the two form $\Omega_{JK}$ to a three form
\f
\Phi_{\mu \nu \gamma }
\ff
which, hence,  is antisymmetric, and which is defined as
\f
\Phi_{0 JK} = \Omega_{JK} , \ \ \ \ \ \ \ \ \ \     d\Phi_{\mu \nu \gamma} =0
\ff
Or we can write:
\f
\Phi_{\mu \nu \sigma} = \eta_{0[\alpha} \Omega_{\beta \gamma ]}= t^\delta \eta_{\delta [\alpha} = \Omega_{\beta \gamma ]}
\ff

We assume there is the structure of a flat metric in the $2+1$ dimensional space; later we could relax this.

We can then define a triple product which generalizes the Poisson bracket
\f
\{ f,g,h\} =( \partial_\alpha f )( \partial_\beta g )( \partial_\gamma h )\Phi^{\alpha  \beta \gamma }
\ff

Let $t^\alpha$ be a time-vector field, and   $s$ a correesponding time coordinate,  so that the derivative of a function $f$ with respect to $s$ is given by
\f
     \frac{d f(q)}{ds} = {\cal L}_t f(q(s)) = t(f)
\ff

By virtue of the above definition, the symplectic potential (\ref{symp0}) is
\f
\Gamma = \int ds  X^\alpha  \Phi_{\alpha \beta \gamma}  \eta^{\gamma \delta}  (\partial_\delta X^\beta )
\ff


\end{enumerate}

\section{Cubic matrix theory as a template for a relational dynamics}

In this section I introduce the cubic matrix models and review what is known about them\cite{cmm,stringloop,EJA,eteralee}.  We will see that they provide an
explicit realization of the ideas we have been discussing.

We start with $N=2P$ dimensional trace free hermitian matrices, $M$, and write the simplest action possible:
\f
S^{basic} = \frac{1}{3} Tr M^3 +  \frac{1}{2m} Tr M^2 
\ff
where $m$ is a dimensionless mass, and we impose
\f
Tr (M)=0
\ff

The equations  of motion are
\f
M^2- \frac{1}{N} Tr M^2= \frac{1}{m} ( M  - \frac{Tr(M)}{N} I   )       = \frac{1}{m} M
\ff

We next reduce this to a matrix theory, of a form previously 
studied in \cite{cmm,stringloop,EJA,eteralee} whose degrees of freedom
are three elements of an algebra:  
\f
M_a \in {\cal A}
\ff
where $a=0,1,2$ and $\cal A$ is some $P \times P$  dimensional matrix algebra (or possibly a more general algebra, possessing a commutator and a trace.)  


We combine these together back into the single tracefree $M$ using the tensor product,
\f
M= \tau^J  \otimes M_J
\label{tftd}
\ff
where $ \tau^{J}$ are the three Pauli matrices, which define also a three-metric,
\f
\eta^{JK} = Tr ( \tau^{J}  \tau^{K})
\ff
They satisfy,
\f
\tau^J \tau^K = \eta^{JK} I + \frac{\imath}{2} \epsilon^{JKL} \tau_L
\ff

The three matrices $M_J$, with $J=0,1,2$ are large $P \times P$  matrices.

The action becomes
\f
S =  \frac{\imath}{2} \epsilon^{abc} Tr ( M_a M_b M_c )  -\frac{1}{2m} Tr ( M_J M_K ) \eta^{JK}  - \lambda^J Tr M_J
\label{Sf}
\ff
The equations of motion are
\f
\epsilon^{abc} [M_b , M_c ]  = \frac{1}{m} M^a  + \lambda^a I
\label{eom}
\ff
and, varying the lagrange multiplier $\lambda^a$
\f
Tr M_J =0
\ff

This theory has a natural triality symmetry given by the permutation group on the three index values, $a,b,c = 0,1,2$.  
The  action is also invariant under the global $SO (1,2)$ ( or $SO(3)$) symmetry group of $\eta_{ab}$:
\f
M_a \rightarrow \Lambda_a^b M_b,
\label{global}
\ff

We will see shortly that  the permutation (or  triality) symmetry represents the freedom to choose some of the degrees of
freedom of each of the matrices as
given rise to frames of reference. Two such choices are related by a permutation that exchanges degrees of freedom
taken to make up the quantum system with those that represent the frame of reference.

On the other hand, the global  symmetries (\ref{global}) will correspond to the usual changes of reference frame under rotations,
boosts etc.

There is also a gauge symmetry,
Let $\cal G$ be the automorphism group of $\cal A$.
Our action has a large gauge symmetry under:
\f
M_a \rightarrow g^{-1} M_a g, \ \ \ \ \ \ \  \ \ g \in {\cal G}.
\label{gauge}
\ff
These will give rise to the unitarity of quantum dynamics, which we will see is represented non-linearly once a reference system is chosen.

\subsection{Solutions and the emergence of dynamics}

We rename two of the matrices:
\f
M_1 =P ; \ \ \ \ \ \ \ \ M_2 =X
\ff

We begin building solutions to the equations of motion of the theory.  

We make extensive use of the compactification of matrixes which produces theories on tori, which are reviewed in \cite{Universal}.

We first imagine for a moment we are in a basis of frequency eigenstates, 
of a quantum particle on a circle, as is reviewed in  \cite{Taylor, Universal}
We chose to diagonalize $X$.  The diagonal
components are $e^{\imath \omega_k}$, where $\omega_k =\frac{ 2\pi k}{N}$.     ie we have
\f
X_{kk} =e^{2 \pi \imath\frac{k}{N}}, \ \ \ \ \  X_{l \neq k} = 0
\label{Xdef}
\ff
So the trace is
\f
TrX = \sum_k  
X_{kk} =\sum_k e^{2 \pi \imath \frac{k}{N}} =0.  
\label{Tr1}
\ff

\subsection{Classical or quantum mechanics?}

As we have set it up so far, the entries of the $N \times N$ matrix, (\ref{Xdef}) are complex numbers, so the sum
(\ref{Tr1}) is the ordinary sum.  If we continue with this we will reach classical dynamics of a one dimensional harmonic
oscillator in a few steps, as we will verify.  But we want to get to quantum mechanics, and to do that we expand
each matrix element $X_{kl}$ and $P_{kl}$ to an $M \times M$ matrix, where we are going to take $M$ to infinity,
along with $P$.  That is we now will have matrices of dimension,
\f
N = 2 P M
\ff
where $P$ and $M$ will both be taken to infinity.  The trace of such a matrix becomes
\f
TrZ = \sum_k  Tr_M X_{kk} 
\ff
that is we take the trace of each $M \times M$ matrix that the diagonal matrix elements $X_{kk}$ indicates, and sum over
them.

We can thus take as an ansatz for the $X$ matrix elements,
\f
X_{kk} =e^{2 \pi \imath\frac{k}{N}} \otimes I_{M\times M}    , \ \ \ \ \  X_{l \neq k} = 0
\label{Xdef2}
\ff

\subsection{Breaking a symmetry to introduce a physical reference frame}

We now introduce the $V$ operator\cite{Universal,Taylor}
which satisfies
\f
[V , X] = diag ( \omega_k e^{2 \pi \imath \frac{k}{N}}) \otimes I_{M\times M}
\ff

Thus we have,
\f
Tr P [V,X] = \sum_k  k  Tr_{M} P(k) X(k)
\label{symp1}
\ff
where $X(k)= X_{kk}$ and $P(k)= P_{kk}$.
By fourier transform we also have, in the limit $N \rightarrow \infty$, 
\f
Tr P [V,X] =\oint ds Tr_{M}  P(s) \frac{dX(s)}{ds} 
\label{symp2}
\ff
where $X(s) = \sum_k e^{\imath 2 \pi s k } X(k)$.

Thus the matrix operator $V$ represents a physical internal clock.  Its action on another degree of freedom, represented by a large
matrix,  is  to extract the time derivative of that degree of freedom.   It is a dynamical frame of reference that gives meaning to the time derivative,
and hence the time  dependence of other degrees of freedom.
In other  words, $V$ is 
an example of an operator that represents a  dynamical, internal reference frame, with a
bounded precision given by $N$.  When we take the limits $N, M \rightarrow \infty$   our matrixes become representations
of quantum operators. In that limit,  $V$ will come to represent a quantum reference frame for time.

We also see by (\ref{symp1},\ref{symp2}) that the introduction of a physical clock, by the specification of a solution, in this way fixes the sympectic structure for the remaining degrees of freedom,
$P$ and $X$.  We will see below that this gives rise to the Heisenberg commutation relations when the limits just mentioned are taken.

We can make any of the three matrix degrees of freedom carry a physical clock, and hence be a physical frame of reference,
by inserting $V$ into the corresponding matrix.   We do this by shifting the matrix by $V$. 

We chose to introduce the frame of reference into the $M_0$ degrees of freedom by a shift 
\f
M_0 = V+ \imath A
\label{shift1}
\ff
where V is fixed and the degrees of freedom are now in the tracefree matrix 
$A$.   

$A$ transforms non-linearly under the symmetries (\ref{global},\ref{gauge}), ie it transforms like a connection. 

The shift (\ref{shift1}) appears to break the symmetries (\ref{global},\ref{gauge}),  but since it leads to the solution of the equations of motion, it can be considered a spontaneous breaking.  We see that the shifted degrees of freedom of $M_0$ take on the role of a connection, related to the emergent notion of time given by the shift of $M_0$ by  $V$.  This is indeed exactly how we
introduce a non-linear realization due to a spontaneous broken symmetry in a Higgs system by shifting the value of the Higgs field.

The action is then, after the shift
\begin{eqnarray}
S&=&  \frac{\imath}{2} \oint ds  Tr_{M} \left (    P(s) [ \frac{dX(s)}{ds} + A(s) X(s) ]   -\frac{1}{2m} ( X(s)^2 + P(s)^2  +  (V+A)^2 ) ) \right ) - 
\nonumber \\
&-&
\lambda^0 Tr A  +\lambda_x Tr X + \lambda_P Tr P
\label{Ss}
\end{eqnarray}
and the equations of motion are
\f
 \frac{dX(s)}{ds} + [ A(s) , X(s) ] = \frac{p}{m}   + \lambda_X I
\ff
\f
 \frac{dP(s)}{ds} - [A(s),  P(s)] = - \frac{x}{m}   + \lambda_P I
\ff
\f
[X(s), P(s)] = \imath \lambda_0 I
\label{UP1}
\ff
Finally, we choose the lagrange multipliers as
\f
A=0 =  \lambda_X = \lambda_P =0
\ff

But at this last step we break the symmetry by keeping $\lambda_0$ finite.    So finally we have,
\f
 \frac{dX(s)}{ds}  = \frac{P}{m}  \ff
\f
 \frac{dP(s)}{ds}  = - \frac{X}{m}  
\ff
\f
[X(s), P(s)] = \imath \lambda_0 I
\label{UP2}
\ff

We note that each of these equations of motion  are $M \times M$ matrix equations, regarded in the limit 
$ M \rightarrow \infty$.  In this limit, and only then, the Heisenberg equations of quantum mechanics are derived
as the limit of  an infinite set of classical equations of motion.  

We should be aware of the apparent paradox that the traces of the left and right sides of the uncertainty relations 
(\ref{UP1},\ref{UP2}) are not equal. For any finite $M$ we would be forced to take $\lambda_0 =0$ and so recover just
$M$ copies of classical mechanics.  Only if we take $M \rightarrow \infty$ can we introduce small terms that go away
as $M \rightarrow \infty$ which makes the expression consistent. 

In the following section we will see how to combine this with choices of physical rulers, which will similarly give rise to quantum reference frames that can anchor definitions of both space and time.  In the next subsection we see how to extend from a particle moving on a circle to one moving in higher dimensional spaces.

\subsection{Introducing additional dimensions.}

Unpacking this let us further\footnote{The contents of this subsection come originally from  \cite{Universal}.} decompose the algebra ${\cal A} $,
\f
{\cal A} = {\cal T}^d \otimes {\cal H}^m_d
\ff
where ${\cal T}^d$ generates translations in $d$ dimensions and ${\cal H}^m_d$ are the Hermitian 
matrices in $M$ dimensions, in the limit $M \rightarrow \infty$.  (We could equivalently  take these latter to be hermitian observables algebra in some Hilbert space.)  Then we  can write
\f
M_1 =   {\cal T}^\mu  \otimes P_\mu (t)  ,    \ \ \ \ M_2 =   {\cal T}_\mu \otimes  X^\mu (t),
\ff
where $X^\mu (t)$ and $ P_\mu (t)$ are each $d$ hermitian matrices  of high dimension $M$.
$M_0$ is as before.
  
$ {\cal T}^\mu$ are translation generators which satisfy
\f
Tr  {\cal T}^\mu =0 , \ \ \ \ Tr  {\cal T}^\mu  {\cal T}^\nu = \eta^{\mu \nu}
\ff
Then the $A$ equations of motion remain 
the Heisenberg algebra, again with $\lambda_0  \sim   \hbar$.
\f
[X^\mu,P_\nu] = \imath \hbar \delta^\mu_\nu I
\ff



The action is now,
\begin{eqnarray}
S&=&  \frac{\imath}{2} \oint ds  Tr_{M} 
(    P_\mu (s) [ \frac{dX^{\mu} (s)}{ds} + A(s) X^\mu (s) ]   -\frac{1}{2m} ( X(s)^2 + P(s)^2 +  (V+A)^2 ) )
\nonumber
\\
&-& 
\lambda^0 (V+A ) +\lambda_{x \mu} X^\mu + \lambda_P^\mu P_\mu
)
\label{Ss}
\end{eqnarray}

and the equations of motion are
\f
 \frac{dX^\mu(s)}{ds} + A(s) X^\mu (s) = \frac{P^\mu }{m}   + \lambda_X I
 \label{eom1}
\ff
\f
 \frac{dP_\nu(s)}{ds} - A(s) P_\nu (s) = - \frac{x_{\nu}}{m}   + \lambda_{P \nu} I
  \label{eom2}
\ff
\f
[X^\mu(s), P_\nu (s)] = \imath \delta^\mu_\nu \lambda_0 I
 \label{eom3a}
\ff
\subsection{Broken duality and the principle of inertia}

The action (\ref{Ss}) and equations of motion (\ref{eom1}-\ref{eom3a}) can be transformed by the following scalings,
and we also set the lagrange multipliers ().
\begin{eqnarray}
X^\mu & \rightarrow & \gamma X^\mu
\nonumber
\\
P_\mu & \rightarrow &  P_\mu
\nonumber
\\
s \  & \rightarrow & \gamma s
\nonumber
\\
\lambda_0  & \rightarrow & \gamma \lambda_0 
\end{eqnarray}

When we take the limit $\gamma \rightarrow 0$,  we find the free equations of motion.
\f
 \frac{dX^\mu(s)}{ds} = \frac{P^\mu }{m}  
 \label{eom4}
\ff
\f
 \frac{dP_\nu(s)}{ds} = 0  \label{eom5}
\ff
\f
[X^\mu(s), P_\nu (s)] = \imath \delta^\mu_\nu \lambda_0 I
\label{eom6}
\ff

We note that the limit $\gamma \rightarrow 0$ breaks the Born duality  that mixes up the $P_\mu$ 
with the $X^\mu$.  Moreover, note that we have {\it derived the principle of inertia}.

We learn a centrally important lesson, which is that in a purely relational
system, the principle of inertia only appears as an emergent law in a limit.   This is because intrinsically, every
degree off freedom is coupled to every other.  The principle of inertia appears only in a limit in which a subset
of degrees of freedom decouple.  Indeed, this is what makes it possible for these nearly decoupled degrees of
freedom to provide a stable frame of reference, against which all the others may be referenced.

To make this completely clear, consider a different limit in which one decouples, not all the $P_\mu$'s, but only
one, or a small number of them.
\begin{eqnarray}
X^\mu & \rightarrow & \gamma X^\mu,  \ \ \ \    \mu = 1,2,3 \   \mbox{otherwise} X^\mu  \rightarrow  X^\mu,
\nonumber
\\
P_\mu & \rightarrow & P_\mu
\nonumber
\\
s \  & \rightarrow & \gamma s
\nonumber
\\
\lambda_0  & \rightarrow & \gamma \lambda_0 
\end{eqnarray}

Note that this is also, essentially,  a statement of Mach's principle.  Acceleration, and its opposite, inertia ,
(the absence of acceleration), arises, when defined in relation to the motions of everything else in the universe.
What is required is only to push almost all the matter up to the distant starts,  decoupling the few near us, 
and so defining the local inertial frames.  This is what the limit $\gamma \rightarrow \infty$ achieves.

\subsection{Recovery of the free relativistic particle}

Is there a way to get the action for a relativistic free particle\footnote{The contents of
this subsection come mainly from \cite{triality}}, from a cubic matrix theory?
There is at least one way which involves extending the Pauli matrices from $3$ to $4$ dimensions. 
This still can be coded by a single tracefree $N \times N$ matrix $\bf M$, with the action still,
\f
S^{2+3 } = Tr M^3  +  Tr  M^2 
\ff
We first expand  the three Pauli matrices to four, 
\f
{\bf M} =\tau^\alpha  \otimes  M_\alpha 
\ff

The  fourth Pauli matrix (corresponding to
$\alpha=3$) proportional
to  $I_{2 \times 2} $ 
with $\alpha = a, 3$, with $\tau^3 = I^{2 \times 2}$.

We now, parametrize the whole $d \times P \times M$ set of four matrices by
\f
M = \tau^1 \otimes \tau^J  \otimes P_J  + \tau^2 \otimes \tau_I \otimes X^I  + \tau^0 \otimes I \otimes (V +A ) 
+ \tau^3 \otimes I \otimes M_3  
\label{1}
 \ff
where 
\f
M_3 = {\cal N}  \otimes I_{M \times M}
\ff
We next introduce again the scaling  parameter $\gamma $
\f
M_0 =    \left ( \frac{2}{\gamma}  \hat{\partial}_t I^\prime + A_0 (t) \right )
\ff
\f
M_1 = \gamma X^\mu (t) {\cal T}_\mu, \ \ \ \ M_2 = P_\mu (t) {\cal T}^\mu
\ff

We take the limit $\gamma \rightarrow 0$, which signals we are breaking Born duality, to find
that the action has become again free,
\f
S^{2+3} \rightarrow \oint dt Tr_M \left ( P_\mu  \frac{d X^\mu }{dt} + A_0 [X^\mu,P_\mu] 
+ {\cal N} \left ( P_\mu P_\nu \eta^{\mu \nu} + {\cal N } ^2 
 \right )
 + \lambda_0 Tr A_0
\right )
\label{SemfH}
\ff
We understand the emergence of the free relativistic partice in the same light as above.
We note that the Lagrange multiplier $\cal N$ gives also the mass, which is arbitrary.
The reader can easily verify that in the limit $ M \rightarrow \infty$ the classical equations of
motion become  the Heisenberg equations for the free relativistic quantum particle.


\subsection{Compactification to Chern-Simons theory}

Once we have understood these elemntary systems, we can go on to do more
sophisticated compactifications, out of which emerge topological field theories,
gauge theories and gravity.

These are described in \cite{Universal}; here we will present only  Chern-Simons theory.   To get it we compactify on a three-torus, $T^3 = (S^1)^3$, go back to the parameterization (\ref{tftd}) and consider only the cubic term,
so that the action is
\f
S^{3}= Tr M^3 = \epsilon^{abc} Tr \left (M_a M_b M_c
\right )
\ff
The equations of motion are simple,

\f
[M_a, M_b ] =0
\label{eom3}
\ff

We compactify on a three torus, using the compactification scheme described in 
\cite{Universal,Taylor}    three times.
We  find a solution to the equations of motion
(\ref{eom3})
\f
M_a = \hat{\partial}_a,  \ \ \ \  \  \  [ \hat{\partial}_a ,  \hat{\partial}_b ] =0
\ff
We expand around this solution
\f
M_a =  \hat{\partial}_a + A_a (x^a )
\ff
to find  Chern-Simons theory for a gauge group, $SU(M)$.
\f
S^f \rightarrow S^{CS} = \int_{T^3} Tr^\prime A \wedge dA + A^3 
\ff

\section{Conclusions}

We developed here two ideas about the structure of physical theories.  The first is that in a background independent theory,  the canonical duality that mixes up configuration with momentum is extended to a triality, that mixes those two
with the temporal reference frame.  Conversely,   the standard canonical form
of dynamics arises from a spontaneous breaking of a symmetry that mixes up
the degrees of freedom of a frame of reference with the other degrees of freedom of the system.

The second idea is that a cubic matrix model allows us to express the first
idea in a general framework that, depending on diferent limits, symmetry breakings and compactifications, can represent classical or quantum mechanics,   or even Chern-Simons theory.

Before closing, I want to mention that
there are a number of older motivations for seeking a unification
of physics in the context of the  cubic matrix models disussed here.

\begin{itemize}

\item{}There are arguments that to make string theory background independent it is necessary to extend it to
a membrane theory\cite{m}.  

\item{}The strategy of starting with a purely cubic background independent action,  whose solutions define a background, appeared earlier in string field theory\cite{SFT}.  These theories were, however, subject to technical issues, which I conjectured could be resolved by framing them as membrane theories 
rather than string theories\cite{cmm,stringloop,EJA}.

\item{}Indeed, one way to express a membrane theory is through a matrix model\cite{CH,DHN,BFSS,IKKT,others,stringloop,EJA}.  
However once one is studying non-linear dynamics for very large or infinite matrices there is a trick which can be used to reduce any non-linear dynamics to quadratic equations.  This is to reduce the degree of equations by introducing auxiliary fields-and then coding these auxiliary fields in the degrees of freedom of an expanded matrix.  You can do this to reduce any non-linear equations to the simplest possible non-linear equations-which are quadratic equations.  Hence any matrix theory should be in its most unified and compressed form when described by a cubic action\cite{Universal}.   This suggests a theory based on an algebra with a triple product defining a generalized trace.  

\item{} In \cite{EJA} the triality of the octonions, which generates a symmetry of the exceptional Joran algebra,  is seen to extend the duality to a triality
symmetry related to the representation theory of $SO(8)$.  This is the unique dimension, $d$, in which the fundamental spinor rep, $S$, (as well as the dual
spinors, $\tilde{S}$ rep ), have each the same dimension as the vector, $V$, namely
$d=8$.  This gives us a cubic invariant from the intertwiner,
\f
{\cal I}: V \otimes S \otimes \tilde{S} \rightarrow C
\ff

Indeed, cubic matrix models have natural triality symmetries\cite{cmm,stringloop,Universal,EJA}.  

  
\item{} The cubic matrix models have been used to propose a unification of gravitational and Yang-Mills dynamics\cite{Universal}.  

\item{}An even deeper unification bringing together the law with the state is described in \cite{stateislaw}.  

\item{}Finally,
quantum mechanics itself can be understood as a consequence of matrix dynamics\cite{hiddenmatrix}-\cite{mefotini-qm};  in these   {\it  relational
hidden varaibles theories}, the be-ables are in fact matrix elements.

\end{itemize}

\section*{Aknowledgements}

We are grateful to, Laurent Freidel,   Thomas Galley, Flaminia Giacomini, and Yigit Yargic for very helpful discussions on drafts of this paper.

This research was supported in part by Perimeter Institute for Theoretical Physics. Research at Perimeter Institute is supported by the Government of Canada through Industry Canada and by the Province of Ontario through the Ministry of Research and Innovation. This research was also partly supported by grants from NSERC and FQXi. We are especially thankful to the John Templeton Foundation for their generous support of this project.


\begin{thebibliography}{99}

\bibitem{TN}Lee Smolin, Temporal naturalism, arXiv:1310.8539, Invited contribution for a special Issue of Studies in History and Philosophy of Modern Physics, on Time and Cosmology, edited by Emily Grosholz.  

\bibitem{3R}Lee Smolin, Three Roads to Quantum Gravity, 2001 Weidenfeld and Nicolson (UK) and Basic Books, (New York)

\bibitem{TR}Lee Smolin,  Time Reborn, Aprill 2013, Houghton Mifflin Harcourt, Random House Canada and Penguin (UK),

\bibitem{SURT}Roberto Mangabeira Unger and Lee Smolin, The Singular Universe and the Reality of Time, Cambridge University Press, November 2014.


\bibitem{QRF0}Giacomini, F., Castro-Ruiz, E. \& Brukner, ?. Quantum mechanics and the covariance of physical laws in quantum reference frames. Nat Commun 10, 494 (2019). https://doi.org/10.1038/s41467-018-08155-0

\bibitem{QRF1}Augustin Vanrietvelde, Philipp A Hoehn, Flaminia Giacomini, Esteban Castro-Ruiz
{\it A change of perspective: switching quantum reference frames via a perspective-neutral framework}
arXiv:1809.00556
doi  10.22331/q-2020-01-27-225

\bibitem{QRF2} Augustin Vanrietvelde, Philipp A Hoehn, Flaminia Giacomini, 
{\it Switching quantum reference frames in the N-body problem and the absence of global relational perspectives},
arXiv:1809.05093 

\bibitem{QRF3}Flaminia Giacomini, Esteban Castro-Ruiz, Caslav Brukner, {\it 
Relativistic quantum reference frames: the operational meaning of spin},
arXiv:1811.08228, doi 10.1103/PhysRevLett.123.090404

\bibitem{QRF4}Esteban Castro-Ruiz, Flaminia Giacomini, Alessio Belenchia, Caslav Brukner, 
{\it Time reference frames and gravitating quantum clocks }
arXiv:1908.10165.

\bibitem{QRF-QG}Philipp A. Hoehn, {\it
Switching Internal Times and a New Perspective on
the Wave Function of the Universe}, Universe 2019, 5, 116; doi:10.3390/universe5050116


\bibitem{cmm}  L Smolin,
{\it M theory as a matrix extension of Chern-Simons theory},
Nucl.Phys. B591 (2000) 227-242,
hep-th/0002009

\bibitem{stringloop}
L Smolin,
{\it The cubic matrix model and a duality between strings and loops},
hep-th/0006137

\bibitem{EJA}Lee Smolin, The exceptional Jordan algebra and the matrix string, hep-th/0104050.

\bibitem{eteralee}Etera Livine and Lee Smolin, BRST quantization of Matrix Chern-Simons Theory, hep-th/0212043.

\bibitem{triality} L. Smolin, Extending dualities to trialities deepens the foundations of dynamics, arXiv: 1503.01424,  International Journal of Theoretical Physics, special issue in memory of David Finkelstein.  10.1007/s10773-016-3168-7

\bibitem{Universal}Lee Smolin, Matrix universality of gauge and gravitational dynamics, 
rXiv:0803.2926.


\bibitem{DoD}L. Smolin, {\it The dynamics of difference,}  arXiv:1712.04799., Foundations of			Physics DOI: 10.1007/s10701-018-0141-8

\bibitem{Taylor}
W Taylor,
{\it D-brane Field Theory on Compact Spaces},
 Phys. Lett. B394,283 (1997),
hep-th/9611042; 
W Taylor,
{\it Lectures on D-branes, Gauge Theory and M(atrices)},
hep-th/9801182; 
W Taylor,
{\it The M(atrix) model of M\_theory},
hep-th/0002016


\bibitem{BFSS}T. Banks, W. Fishler, S. H. Shenker, L. Susskind, M theory as a matrix model: a conjecture hep-th/9610043; Phys. Rev. D55 (1997) 5112.



\bibitem{compactification}Washington Taylor,  {\it Lectures on D-branes, Gauge Theory and M(atrices)}, arXiv:hep-th/9801182; 
{\it The M(atrix) model of M-theory}, arXiv:hep-th/0002016. 

\bibitem{harold2}H. Grosse, H.  Steinacker, {\it Finite Gauge Theory on Fuzzy $CP^2$}, arXiv:hep-th/0407089,  Nucl.Phys. B707 (2005) 145-198. 

\bibitem{m}C. M. Hull and P. K. Townsend, Unity of superstring dualities, Nucl. Phys. B348 (1995) 109; E. Witten, String theory in various dimensions, Nucl. Phys. B443 (1995) 85; P. Townsend, (M)embrane theory on T 9, Nucl. Phys. (Proc. Suppl) 68 (1998) 11-16; hep- th/9507048, in Particles, Strings and Cosmology, ed. J. Bagger et al (World Scien- tific,1996); hep/9612121; I. Bars, hep-th/9608061; hep-th/9607122; Petr Horava, M- Theory as a Holographic Field Theory, hep-th/9712130, Phys.Rev. D59 (1999) 046004.

\bibitem{CH}M. Claudson and M. Halpern, Nucl. Phys. B250 (1985) 689.

\bibitem{DHN}B. DeWitt, J. Hoppe, H. Nicolai, Nuclear Physics B305 (1988) 545.
 
\bibitem{IKKT}N. Ishibashi, H. Kawai, Y. Kitazawa and A. Tsuchiya, A large N reduced model as superstring hep-th/9612115; Nucl. Phys. B498 (1997) 467; M. Fukuma, H. Kawai, Y. Kitazawa and A. Tsuchiya, String field theory from IIB matrix model hep-th/9705128, Nucl. Phys. B (Proc. Suppl) 68 (1998) 153.
[10] L. Smolin, Covariant quantization of membrane dynamics, hep-th/9710191, Phys.Rev. D57 (1998) 6216-6223.

\bibitem{others}C. M. Hull and P. K. Townsend, Unity of superstring dualities, Nucl. Phys. B348 (1995) 109; E. Witten, String theory in various dimensions, Nucl. Phys. B443 (1995) 85; P. Townsend, (M)embrane theory on T 9, Nucl. Phys. (Proc. Suppl) 68 (1998) 11-16; hep- th/9507048, in Particles, Strings and Cosmology, ed. J. Bagger et al (World Scien- tific,1996); hep/9612121; I. Bars, hep-th/9608061; hep-th/9607122; Petr Horava, M- Theory as a Holographic Field Theory, hep-th/9712130, Phys.Rev. D59 (1999) 046004.


\bibitem{laurent-string}Laurent Freidel, Robert G. Leigh, Djordje Minic, {\it  Born Reciprocity in String Theory and the Nature of Spacetime}, arXiv:1307.7080; {\it   Quantum Gravity, Dynamical Phase Space and String Theory}, arXiv:1405.3949; {\it Metastring Theory and Modular Space-time}, arXiv:1502.08005 

\bibitem{nfs}P.Zanardi and M.Rasetti, Phys.Rev.Lett.79 3306 (1997);
D. A. Lidar, I. L. Chuang and K. B. Whaley, Phys. Rev. Lett. 81, 2594 (1998)
[arXiv:quant-ph/9807004];
E.Knill, R.Laflamme and L.Viola, Phys.Rev.Lett. 84 2525 (2000);
J. Kempe, D. Bacon, D. A. Lidar and K. B. Whaley, Phys. Rev. A 63, 042307 (2001).

\bibitem{nfs-gauge}T. Konopka and F, Markopoulou, {\it Constrained Mechanics and Noiseless Subsystems}, arXiv:gr-qc/0601028. 

\bibitem{fot-dav}D. W. Kribs and F. Markopoulou, arXiv:gr-qc/0510052;
 F. Markopoulou, {\it Towards Gravity from the Quantum},   arXiv:hep-th/0604120.  
  




\bibitem{Maldacena}Juan M. Maldacena,  {\it  The Large N Limit of Superconformal Field Theories and Supergravity}, arXiv:hep-th/9711200,  Adv.Theor.Math.Phys.2:231-252,1998.

\bibitem{stateislaw}Lee Smolin, Unification of the state with the dynamical law, arXiv:1201.2632.  Foundations of 	Physics: Volume 45, Issue 1 (2015), Page 1-10 DOI 10.1007/s10701-014-9855-4

\bibitem{hiddenmatrix}Lee Smolin, {\it Matrix models as non-local hidden variables theories}, hep-th/0201031.

\bibitem{adler}S. Alder,  {\bf Quantum theory as an emergent phenomenon}, 2004 -Cambridge University Press New York; 
{\it Statistical dynamics
of global unitary invariant matrix models as pre-quantum
mechanics}, hep-th/0206120.

\bibitem{artem}A.  Starodubtsev, {\it A note on quantization of matrix models},  arXiv:hep-th/0206097,  Nucl.Phys. B674 (2003) 533-552. 


\bibitem{mefotini-qm}F. Markopoulou and L. Smolin, {\it Quantum Theory from
Quantum Gravity}, gr-qc/0311059, Phys.Rev. D70 (2004)
124029.

\bibitem{SFT}Gary T. Horowitz, Joseph Lykken, Ryan Rohm, and Andrew Strominger, {\it Purely Cubic Action for String Field Theory},  Phys. Rev. Lett. 57, 283.

\bibitem{matrixstring}Lubos Motl, {\it Proposals on nonperturbative superstring interactions}, hepth/9701025; Tom Banks, Nathan Seiberg, {\it Strings from Matrices}, Nucl.Phys. B497
(1997) 41-55,h ep-th/9702187;  R. Dijkgraaf, E. Verlinde, H. Verlinde, {\it Matrix String Theory,}
Nucl.Phys. B500 (1997) 43-61, hep-th/9703030




\bibitem{janetal}J. Ambj¿rn, Y.M. Makeenko, J. Nishimura and R.J. Szabo, {\it  Lattice Gauge Fields and Discrete Noncommutative Yang-Mills Theory},  arXiv:hep-th/0004147. 



\bibitem{qgraphity}T. Konopka, F. Markopoulou, S. Severini {\it Quantum Graphity: a model of emergent locality}, arXiv:0801.0861. 

\bibitem{phys-comp}For example, see S. Lloyd, {\it  Universe as quantum computer}, arXiv:quant-ph/9912088, 
 Complexity 3(1), 32-35 (1997); {\it Computational capacity of the universe}, arXiv:quant-ph/0110141, Phys.Rev.Lett. 88 (2002) 237901. 


\bibitem{cubicmatrix}L. Smolin,  {\it M theory as a matrix extension of Chern-Simons theory}, arXiv:hep-th/0002009,  Nucl.Phys. B591 (2000) 227-242;  {\it The cubic matrix model and a duality between strings and loops},  arXiv:hep-th/0006137; {\it The exceptional Jordan algebra and the matrix string}, arXiv:hep-th/0104050; 
E. R. Livine, L. Smolin, {\it  BRST quantization of Matrix Chern-Simons Theory}, arXiv:hep-th/0212043.   







\end{thebibliography}
\end{document}